\documentclass[]{article} 
\usepackage[margin=0.75in]{geometry}
\usepackage{graphicx}
\usepackage{amsmath}
\usepackage{hyperref}
\usepackage[table,x11names]{xcolor}

\usepackage{fancyhdr}

\newcommand{\footremember}[2]{%
	\footnote{#2}
	\newcounter{#1}
	\setcounter{#1}{\value{footnote}}%
}
\newcommand{\footrecall}[1]{%
	\footnotemark[\value{#1}]%
} 

\title{An analysis of gamma-ray data collected at traffic intersections in Northern Virginia}
\author{%
	Nathan Hoteling\footremember{RSL}{Remote Sensing Laboratory} , 
	Eric T. Moore~\footnote{Special Technologies Laboratory: mooreet@nv.doe.gov},
	William P. Ford~\footrecall{RSL}~\footnote{The Probitas Project: wford@probitas-project.com}, \\ 
	Thomas McCullough\footnote{Tenica and Associates, LLC}, 
	Lance McLean\footnote{Noblis, Inc.}}

\begin{document}

\maketitle
\thispagestyle{fancy}
\begin{abstract}
Gamma-ray spectral data were collected from sensors mounted to traffic signals around Northern Virginia.  The data were collected over a span of approximately fifteen months.  A subset of the data were analyzed manually and subsequently used to train machine-learning models to facilitate the evaluation of the remaining ~50k anomalous events identified in the dataset.  We describe the analysis approach used here and discuss the results in terms of radioisotope classes and frequency patterns over day-of-week and time-of-day spans.   Data from this work has been archived and is available for future and ongoing research applications.
\end{abstract}

\section{Introduction}

\subsection{The radiation background}

Ambient background radiation consists of the terrestrial environment, cosmic radiation, and a variety of anthropogenic sources~\cite{UNEP16}.  The terrestrial environment is generally comprised of varying quantities of Potassium (K), Uranium (U), and Thorium (Th), present in rocks, soil and other materials.  Among these sources, U and Th daughter products are also radioactive and give rise to airborne radiation like radon and its daughters.  In fact, one of the most prominent fluctuations in static radiation measurements comes from changes in these airborne products primarily as a result of weather conditions.  For example, precipitation events can scavenge these products from the atmosphere, pressure-shifts and changes in soil moisture can change the rate of radon diffusion from the soil, and wind conditions can affect how airborne products concentrate or disperse~\cite{ICRU53}.  Cosmic radiation is a relatively small component from high-energy particles entering Earth’s atmosphere from space.  The magnitude of this component grows somewhat with elevation since the amount of atmosphere is reduced at higher altitudes.  A relatively small component of anthropogenic radiation is derived from historic nuclear weapons tests and nuclear accidents, however a much larger component is from medical exposures.  In fact, from the 1980’s to the early 2000’s the growth in medical radiation exposures led to an increase in the estimated annual dose in the US from 3.6 to 6.2 mSv/y~\cite{NCRP93, NCRP160}.  Another relatively small component of the anthropogenic radiation background comes from industrial radiation sources.  For example, instruments using radiological materials are commonly used in construction for moisture, density, and radiography measurements~\cite{Jackson04,NAP08}.

Medical and industrial sources are unique within the radiation background because they are often manifested as discrete sources and not necessarily fixed or distributed.  This also places these sources into a category of so-called nuisance sources when considered within the context of radiation monitoring for security applications~\cite{Kouzes11}.  

\subsection{Machine-learning and radionuclide identification}

The primary modality for identifying radioactive materials in the field is through the measurement of gamma radiation.  The gamma-ray energy spectrum is a unique fingerprint that facilitates the identification of most radionuclides typically encountered in the field.  As such, the problem of radionuclide identification would seem to be well suited for machine-learning applications.  In fact, there have been a number of investigations into the possibility of using various machine-learning methods for radionuclide identifications~\cite{OLMOS1992167,Portnoy04,Sharma12,Gerrit17,Kamuda2017,Masala19,Durbins2020}. 
Most of these have been focused on nuclear security and safeguards applications.  The authors of this work and colleagues have studied similar applications with modeled data~\cite{ford2019threat,moore2019application,Hague_2019} as well as with a subset of the data presented here~\cite{moore2020transfer}.  

\subsection{What’s out there?}

Historically, efforts to characterize the radiation background have focused primarily on environmental measurements, and characterizations of nuisance sources have typically started from medical and/or industrial source usage statistics~\cite{Kouzes11}.  Each of these methods yields valuable and insightful information, however, neither produce a complete picture of what one might actually encounter in the field.  For example, the ability to detect sources used in medical and industrial applications requires that they are detectable from some monitoring position.  A medical source used only within the medical facility is not likely to be detected by a static portal monitor, but a radiopharmaceutical injected into a patient who then travels home along a public roadway would be detectable for some period of time.  Hence, to obtain a more complete understanding of “what’s out there” the work described here takes the novel approach of measuring radiation anomalies in the urban and suburban environment over the course of approximately fifteen months.  A portion of the resulting data are used to develop a model approach based on machine-learning algorithms and then applied to the remaining data to facilitate the characterization of radiation anomalies specifically at locations in Northern Virginia.  The data gathered in this work is documented and stored so as to facilitate future work.

\section{NOVArray: The Northern Virginia Array}

NOVArray detectors used in this analysis are each comprised of a 2”x4”x16” NaI gamma-ray sensor mounted to an Ortec Digibase and outfitted with a telemetry system.  Data from the sensors are sent via the cell network to a database owned and operated by Two Six Labs.  The data are recorded with a frequency of 1 Hz and spectra are stored in 1024 channels spanning approximately 30 keV to 3 MeV in energy.  

\begin{figure}
	\centering
	\includegraphics[width=2in]{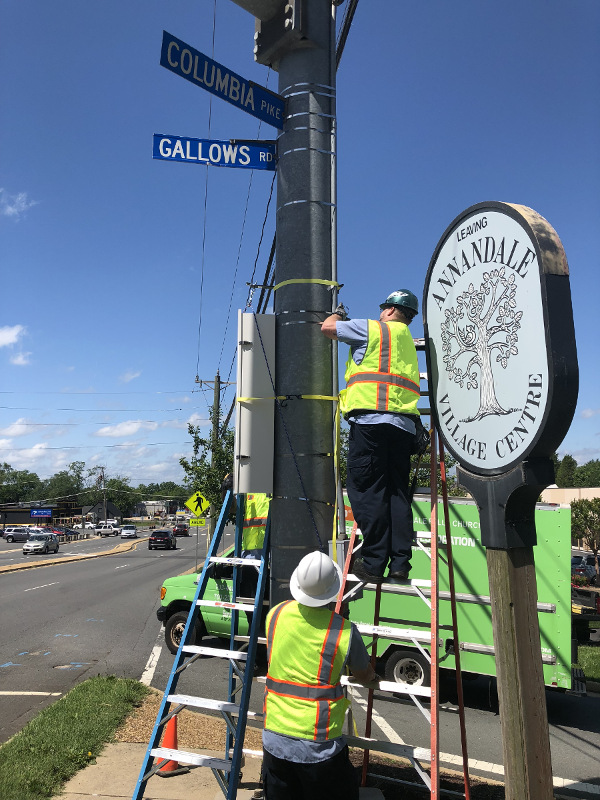}
	\caption{A typical NOVArray sensor configuration.  The detector is housed in the white box and mounted to a traffic signal post.}
	\label{fig:installation}
\end{figure}

The full array included nineteen sensors mounted to traffic signals at locations throughout Northern Virginia.  The sensors are mounted to upright posts supporting traffic lights and oriented as best as possible to the point of closest approach to traffic in the intersection served by the signal.  The sensors are connected to the power grid via 110 VAC GFI outlets and housed in a weather resistant outer enclosure.  The image in Figure~\ref{fig:installation} shows an example of a sensor being installed.

The sites for NOVArray sensors were chosen to maximize the probability that a vehicle traveling through the area would encounter a sensor. This was determined qualitatively by studying the open-source traffic patterns available from the Commonwealth of Virginia~\cite{VA2020}. With this information, it was possible to identify the primary roadways in the region and select locations to specifically capture as much of the traffic as possible while attempting to minimize redundancy between sensors.
NOVArray data were collected from 2018-07-15 to 2019-10-01, a total of 443 days. Over the course of this time period, the anomaly detection algorithm applied here marked 86,165 anomalies. However, as will be discussed later, some of these arose due to instrument issues and were subsequently removed, so a more accurate count of 48,974 anomalies is more appropriate.

\section{Analysis Methodology}

\subsection{Anomaly Detection} 

To identify regions of interest within the streaming dataset, an anomaly detection algorithm was applied to the spectral data.  For this work, the Information Content Anomaly Detection (ICAD) algorithm was used.  The details of this approach are beyond the scope of this report.  In summary, the method uses Shannon entropy as a metric to describe the quality of a template fit, where the template is the background spectrum for a given sensor~\cite{Anom2020}.

\subsection{Manual Labeling} 

\begin{figure}
	\centering
	\hspace*{-.25in}
	\includegraphics[width=5.in]{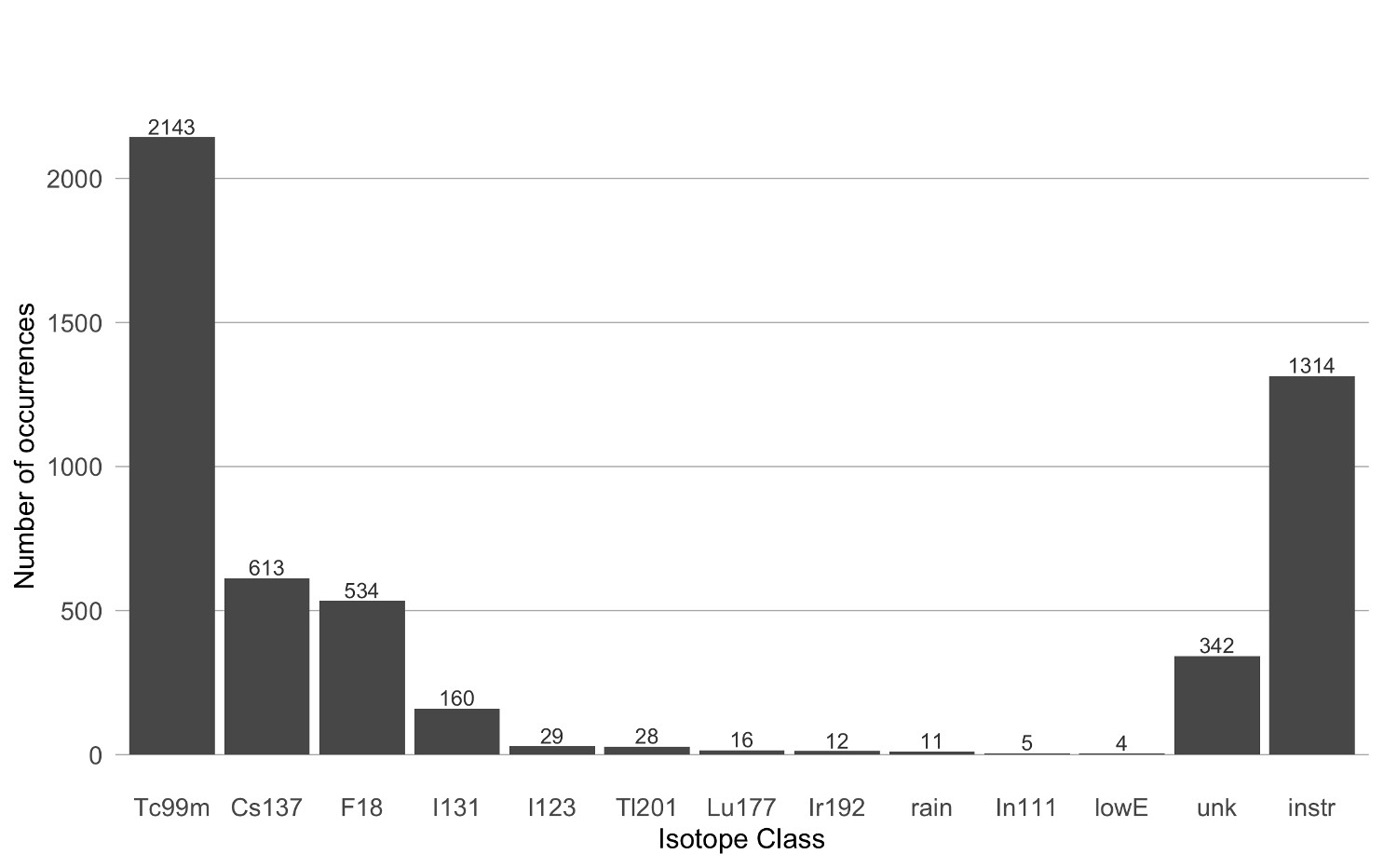}
	\caption{Distribution of isotope class assignments derived from a manual analysis of the first 30 Days of data collected with NOVArray.}
	\label{fig:isotopeDistribution}
\end{figure}

Anomalies recorded over the first 30 days of data collection were reviewed manually and assigned isotope labels.  This brute force analysis included a comprehensive review of 7,967 events by one analyst, and a complementary review by a second analyst.  Labels were compared between analysts and consistency checks were carried out in a qualitative fashion using linear and logistic regression models.  Any disagreement between these approaches triggered a tertiary inspection and a decision was made to revise the label or discard the anomaly.  Using this methodology, a final subset consisting of 3,555 anomalies was identified for use in developing a model to automatically label events in the remaining data.  The distribution of isotope classes is illustrated in Figure~\ref{fig:isotopeDistribution}.

The most significant outcome of the manual review process was a robust collection of labeled data.  However, the detailed inspection necessary to achieve this also helped facilitate the discovery of a number of features in the dataset.  For example, around 40\% of the anomalies identified in the first 30 days were characterized by a sudden drop in count rate, indicating the presence of an instrument error, not a real, substantive radiological signal.  The feature appeared to be mostly random, however some sensors seemed to be affected more than others.  Since the feature could be readily distinguished by a sudden drop in counts, it was straightforward to identify and remove them with an unsupervised approach like kmeans clustering.  With this method, ten unique clusters were characterized, and new events beyond the first 30 days were marked by comparing the Euclidean distance between this profile with that of each cluster.

\begin{figure}
	\centering
	\hspace*{-.75in}
	\includegraphics[width=6.5in]{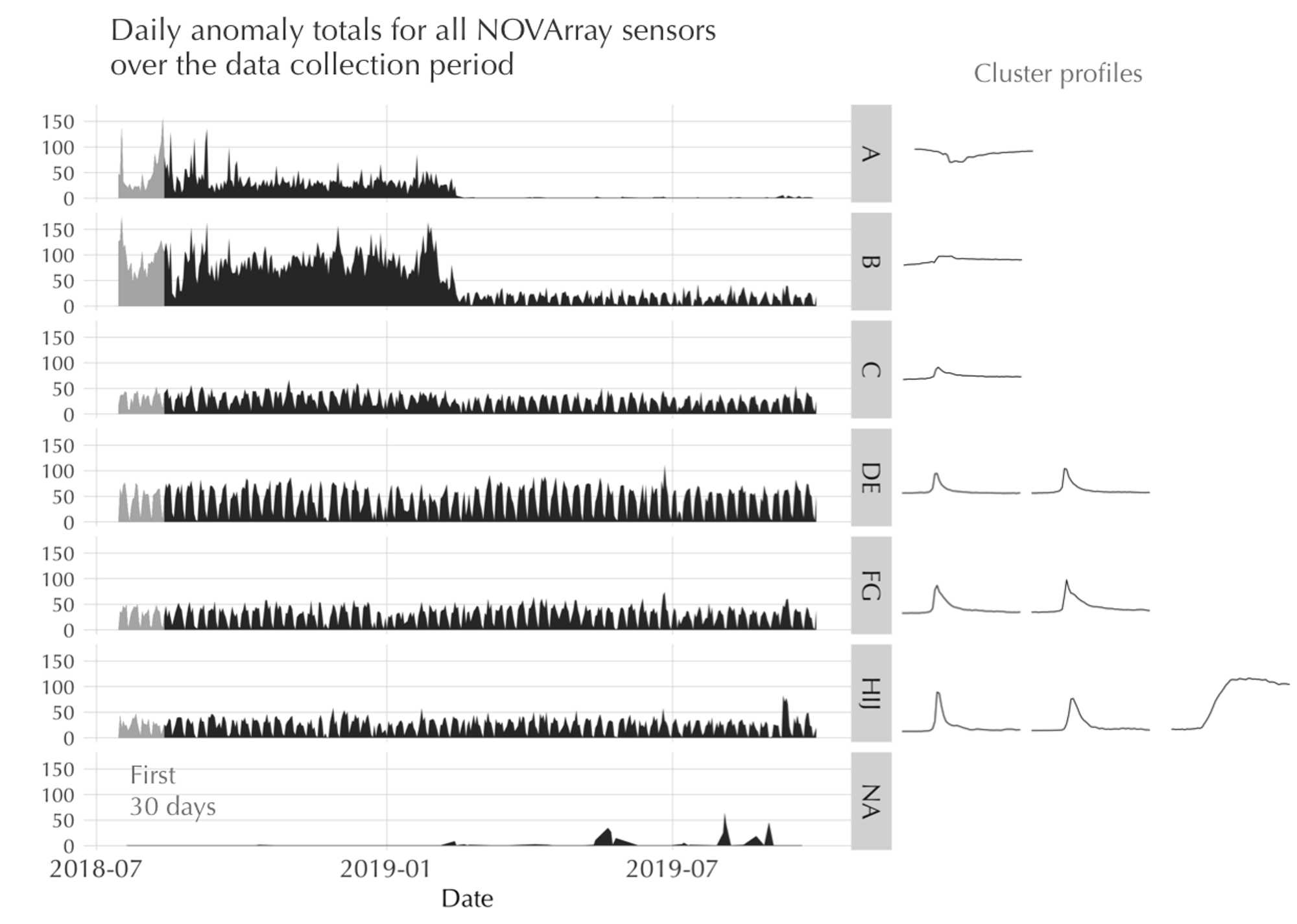}
	\caption{Daily anomaly count for each time series cluster over the course of the data collection period. The first 30 days of data collection are depicted in light grey.  The general shape of each cluster is presented at right.  Note the sawtooth pattern exhibited in the plot, which comes from the fact that fewer source encounters are seen on weekends.  On close inspection, one can also spot increases in ‘DE’, ‘FG’, and ‘HIJ’ groups from the ground-truth measurement campaigns discussed later in this paper.}
	\label{fig:DailyAnomaly}
\end{figure}

Eight months into the data collection period, firmware upgrades applied to each sensor in the field not only eliminated these features, but also removed significant noise from the count rate profile.  After this, it became clear that many of the weaker anomalies were also a result of this instrument issue, as can be seen in Figure~\ref{fig:DailyAnomaly}.  Here, the time series clusters are labeled by letters and generally ordered according to the signal intensity. The oscillating pattern evident in these plots illustrates the dramatic contrast in anomalies seen during weekdays as compared to those seen on weekends.  This will be discussed in more detail later.  Also, a noticeable jump in anomalies for DE, FG, and HIJ groupings can be seen in June and September, corresponding to the two ground-truth test campaigns.

\begin{figure}
	\centering
	\hspace*{-.1in}
	\includegraphics[width=5.in]{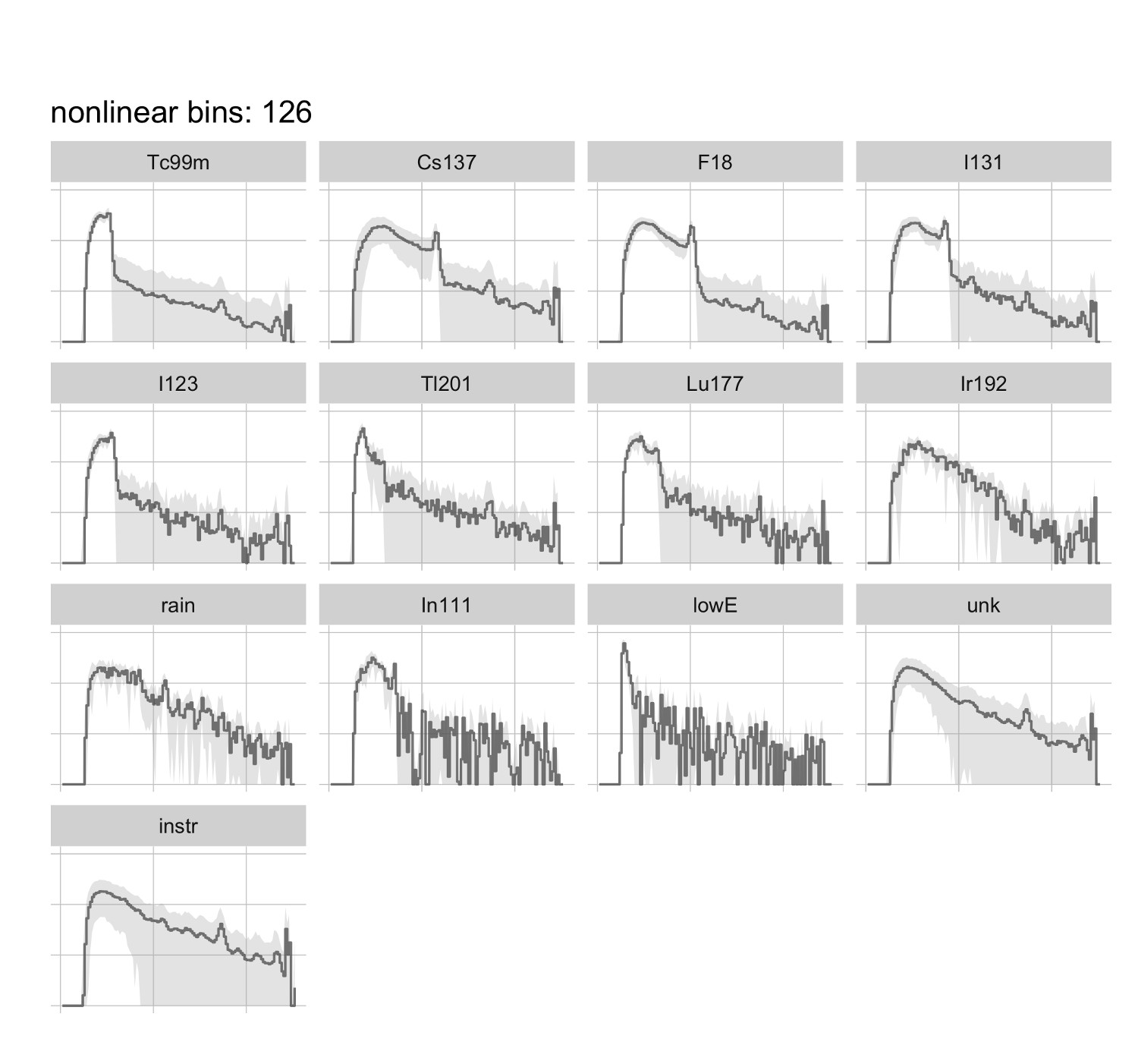}
	\caption{The gamma-ray spectra associated with each isotope class identified in an analysis of the first 30 days of NOVArray data.  The primary line in each plot shows the mean value for the normalized spectra and the shaded regions depict the standard deviation.  Note that some of the histograms show clear, well-defined shape and some are more ambiguous. Note histograms are plotted on a log scale.}
	\label{fig:spectra}
\end{figure}

Isotope labels for each anomaly were assigned on the basis of spectral shape, since every gamma-emitting radioisotope has a unique energy spectrum.  However, the exact shape can appear to vary somewhat based on signal strength and in the presence of shielding materials.  For example, since gamma rays are attenuated more effectively at low energy, this can have a significant impact on how a spectrum measured in the field compares with that from a bare source or a computed spectrum.  To illustrate this effect, we show the mean histograms measured for each isotope class in~\ref{fig:spectra}.  In particular, we note the relatively ambiguous shape exhibited by the Ir-192 class.  This isotope is routinely used in industrial radiography applications with high-activity sources~\cite{NAP08}, so signals measured with NOVArray are likely from relatively strong sources contained inside of well-shielded transport containers.  The result is that the normally well-defined spectral shape becomes somewhat ambiguous since the gamma rays scatter as they pass through the dense shielding material.  A similar effect could be seen for some of the Cs-137 anomalies, since these are often packaged into devices used by construction companies for measuring density.  This fact may account for why some other common radiography sources like Co-60 do not appear to be present in the data.  In contrast, radioisotopes used for medical procedures are often injected into the patient and therefore human tissue is the most significant shielding material between the source and the detector.  These radioisotopes tended to largely maintain their spectral shape.

Three of the isotope labels do not correspond to a unique radioisotope.  The first, “rain” represents the sudden increase in ambient radiation levels with the onset of a precipitation event.  This is largely due to the scavenging of radon decay products in the atmosphere during a precipitation event~\cite{ICRU53}.  When this occurs the gross count rates tend to increase systematically, and the energy spectrum begins to show some distinct characteristics, as can be seen in Figure~\ref{fig:spectra}.  The second, “lowE”, may be consistent with an x-ray source owing to a sharp low energy photopeak, or it could be associated with another radioisotope with a primarily low-energy signal.  The true source of this signal is not presently known to these authors.  Finally, we use the F-18 isotope class somewhat generically to capture any positron-emitting radioisotope, as marked by a distinct photopeak at 511 keV. While F-18 is the most commonly-used radioisotope used in PET scanning, it is certainly possible the some of these events correspond to other medical isotopes like Cu-64 or C-11 ~\cite{Duncan98}.

\subsection{Model-based Labels}

Results from the initial 30 days were used to develop a model that could efficiently and reliably determine labels for the remaining events in the dataset.  The model development process generally consisted of three distinct phases: 1) accounting for significant class imbalance, 2) optimization of the bin structure, and 3) algorithm training.  Each of these is described in detail below.

\subsubsection{Class imbalance} 

There is a significant imbalance between isotope classes in the data, with Tc-99m consisting of about 60\% of all anomalies, and the combination of I-123, Tl-201, Lu-177, Ir-192, rain, In-111, and lowE making up less than 3\%.  This poses a problem for model training since the result will be biased toward the more prominent isotope classes~\cite{unbalanced, Johnson19}.  One way to address this is to sample from existing data to create a more even distribution of classes to train on.  To do this, we created a random split for each isotope class, in which each class was split into two equal-sized chunks to be used separately for training and testing.  Simulated events were then created by first selecting a random event from the respective chunk, and then binwise Poisson sampling was used to generate a “new” histogram.  This process was repeated until there was a total of 300 samples for each isotope class for both training and testing splits.  

The procedure described above was repeated five times with different splits to accommodate the possibility that one problematic event might have a disproportionate effect on the results.

\subsubsection{Bin structure}

The energy spectra for NOVArray data are recorded with 1024 bins, however since the measured signals tended to be fairly weak, the unique shape characteristic of each isotope class frequently appeared ambiguous until these bins were merged and the binwise counts correspondingly increased.  In fact, NaI spectra are frequently collected over 256 bins; not only does this improve the binwise statistics, it also results in significant improvements in the computation time required for analysis.  

Since photopeak resolution for NaI detectors varies as a function of gamma ray energy, it is also sensible to apply a nonlinear binning scheme in which the bin width varies as a function of energy resolution.  To test this, we compared linear and nonlinear binning schemes for several prospective models.  The results consistently yielded improvements in model accuracy by up to several percentage points for the nonlinear binning schemes.

\subsubsection{Model configurations}

Five different algorithms were investigated here: Linear Regression (GSN), Logistic Regression (BIN), Support Vector Machine (SVM), Random Forest (RFT), and Linear Discriminant Analysis (LDA).  Each was trained, or fit, with the training set described above consisting of 300 individual histograms for each isotope class using implementations available in R~\cite{Rproject}.  The resulting models were likewise tested against 300 histograms for each isotope class.  Each was tested with both linear and nonlinear binning structures, and with several different numbers of bins.  The top ten performing model configurations are listed in Table~\ref{tab:modelPerf}.  Here, the accuracy indicates the overall percent of anomalies that were assigned correct labels, averaged over five training/testing splits, and sigma is the standard deviation.   With the exception of the Random Forest, all of the models yielded around 80\% accuracy under the 64-bin nonlinear configuration.

\begin{table}\centering
\begin{tabular}{c c c c c c}
	\hline
	& Algorithm & Structure &	bins &	Accuracy & $\sigma$ \\
	\hline \hline 
\rowcolor{lightgray}	1 & LDA & non & 64  & 82.0 & $\pm$ 3.5  \\
\rowcolor{lightgray}	2 & GSN & non & 64 & 81.8 & $\pm$  2.7 \\
	3 & LDA & lin & 64 & 81.0 & $\pm$  3.3 \\
\rowcolor{lightgray}	4 & SVM  & non & 64 & 80.9 & $\pm$  5.1 \\
	5 & BIN & non & 128 & 80.8 & $\pm$  6.2 \\
\rowcolor{lightgray}	6 & BIN & non & 64 & 80.2 & $\pm$ 	2.9 \\
	7 & LDA & non & 128 & 79.1 & $\pm$  7.3 \\
	8 & LDA & lin & 128 & 78.8 & $\pm$  4.3 \\
	9 & BIN & lin & 64 & 78.6 & $\pm$  2.2 \\
	10 & GSN & non & 128 & 78.6 & $\pm$ 6.1 \\
	\hline
\end{tabular}
\caption{The top ten performing model configurations from the testing phase are listed here.  Each model configuration includes a unique combination of algorithm, bin structure, and number of bins.  See text for details.}
\label{tab:modelPerf}
\end{table}

\begin{figure}
	\centering
	\hspace*{-.5in}
	\includegraphics[width=5.5in]{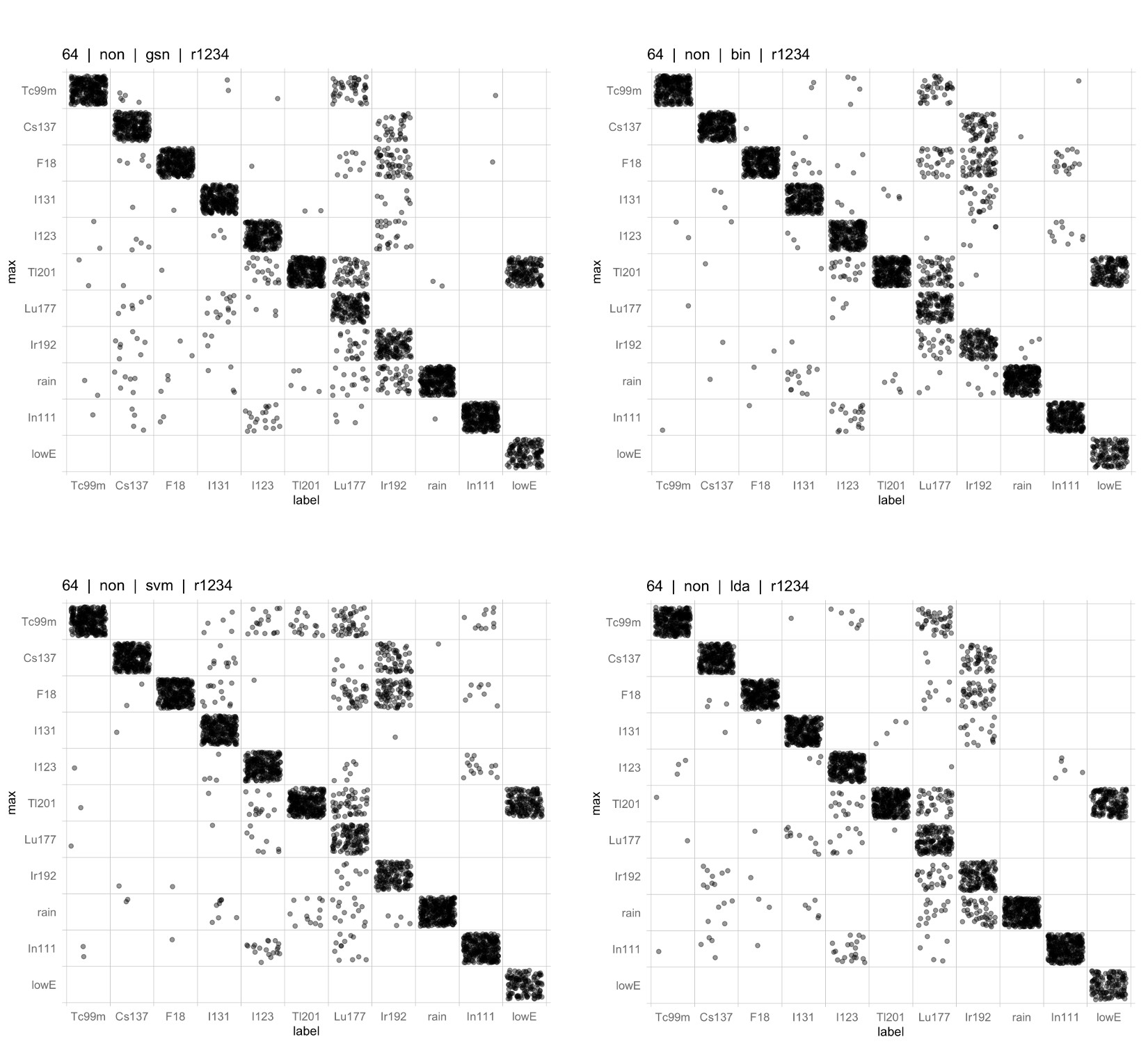}
	\caption{Confusion matrices for a selection of the top performing model configurations from this investigation.  The character string at the top of each plot indicates the bins, bin structure, model, and random number seed used to split test and training sets.}
	\label{fig:conf_mxts}
\end{figure}

To investigate the performance of each model configuration as a function of isotope class, the confusion matrices for four of the top-performing configurations are presented in Figure~\ref{fig:conf_mxts}.  Here, one can see that performance is weaker in particular for Lu-177, Ir-192, and lowE isotope classes.  This is not entirely unexpected since those isotope classes were among the sparsest in the dataset, with 16, 12, and 4 natural occurrences, respectively.  However, some of the other rare isotope classes, like Tl-201 and rain appear to perform reasonably well.  While there generally aren’t significant differences in the detailed performance between models, a couple of features are worth noting: first, SVM and BIN appear to perform somewhat better than GSN and LDA for Cs-137, however the opposite is true for lowE.  The SVM model configurations performed poorly for In-111, but BIN performed well for this isotope class.

Looking even deeper, one can study the isotope probability values assigned to each isotope class for each model.  What is of interest here is whether, for incorrect labels, the model gave less confident assignments or, more importantly, whether the model gives an elevated probability for the correct isotope class even if the final assignment is incorrect.  For example, if the model yielded a 60\% probability for an incorrect label and a 40\% probability for the correct label, then the assignment depicted in Figure~\ref{fig:conf_mxts} would be incorrect, but the result cannot be considered completely wrong.  In fact, this was precisely the case for many of the results, so it is feasible that the outcome can be improved upon.  For this reason, instead of selecting a single “winning” model, we use an ensemble of the four models depicted in Figure~\ref{fig:conf_mxts}.  The ensemble combines the isotope-specific probabilities from GSN, BIN, SVM, and LDA models into a Random Forest.  The result, cross-validated against a small sample of held-out data, performed exceptionally well, however a precise performance metric is not reported here because the held-out data were not completely independent from the dataset used to train the Random Forest.  This is because the simulated data were sampled from histograms in both datasets; previous investigations showed that this leads to inflated performance estimates.  Instead, the ensemble model will be judged by comparing the results to “ground truth” data collected over a series of test campaigns carried out on the array.

\section{Results} 

Before applying the ensemble model described in the sections above to the remaining anomalies identified in the dataset, several additional events needed to be removed due instrument issues encountered at various points over the course of the 15-month data collection period.  As was discussed above, time series profiles were used to identify anomalies characterized by a sudden drop in count rate (cluster A in Figure~\ref{fig:DailyAnomaly}).  The same method was used to identify cluster B anomalies; those encountered prior to firmware upgrades on 14-February-2019 are removed from the analysis since most of these were likely not associated with a real radiological signal.  In addition to these, some sensors lost calibration over the course of the data collection period, usually associated with an instrument outage.  Once these events were removed, a total of 48,974 anomalies remained for analysis (note that this value does not consider those anomalies identified in the first 30 days).

\begin{figure}
	\centering
	\hspace*{-.75in}
	\includegraphics[width=6.in]{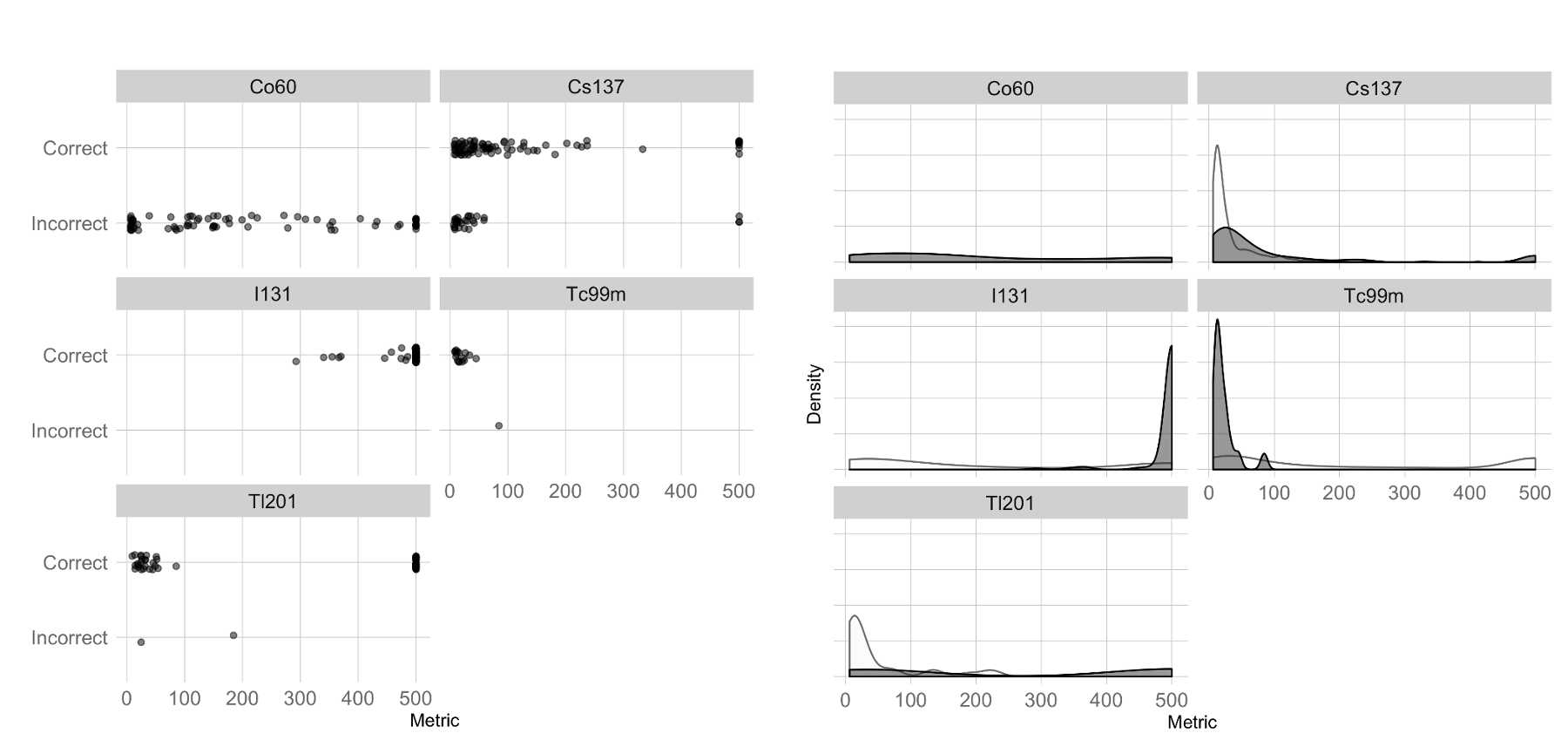}
	\caption{Results from the ensemble model applied to ground truth measurements.  The plot at left shows the number of correct and incorrect labels as a function of anomaly metric value.  At right are density plots showing the alarm metric values for controlled measurements (shaded) and from the manual labels applied to data collected over the first 30 days. Note, the ground truth measurements included Co-60, however, there were very few Co-60 events in the field data and due to this neglected in the previous discussion.}
	\label{fig:ensembleGroundTruth}
\end{figure}

\subsection{Ground truth measurements}

Two measurement campaigns were carried out over the course of the data collection period during which time radioactive sources were driven past a selection of the sensors several times throughout the day.  Since the primary purpose of these test campaigns was to serve as a test for the model predictions, we list the results in the left panels of Figure~\ref{fig:ensembleGroundTruth}.  However, when interpreting these results, it is important to consider how well the source pass-by measurements fit into the applicability domain of the ensemble model~\cite{appDomainQSAR}.  For example, we note that in some cases the pass-by measurements show a different distribution of alarm metric values than what was present in the initial 30 days dataset used for training, as can be seen from the density plots depicted in the right-hand panels of Figure~\ref{fig:ensembleGroundTruth}.  Here, a high value for the alarm metric indicates that the spectrum shape is likely quite clear, and a low value means that it is likely to be somewhat ambiguous and therefore more difficult to distinguish.  The I-131 test campaign data gave very high alarm metric values, whereas the field data were more evenly distributed.  For this reason, the fact that the model yielded perfect results for this isotope class should be interpreted with care.  On the other hand, the Cs-137 alarm metrics are fairly similar between test campaign and field data.  Hence, the fact that the model had some difficulty for the very weak events is a meaningful result.  Interestingly, the Tc-99m test campaign data tended to be weaker than the field data, so the fact that the model performed so well for this isotope class is also a meaningful result.  

It is also worth noting that the model gave an incorrect assignment for Co-60 in every case.  This is because Co-60 was not identified in the first 30 days and, as such, the model does not include an isotope class for this radionuclide, another artifact of the limited applicability domain for the ensemble model.  A closer inspection indicated that these events tended to be assigned to the “rain” isotope class.  This is a reminder that any additional rare signals that may be present in the full dataset but not in the first 30 days will also have incorrect assignments.  

In general, the test campaign results indicate that the model predictions are quite good, and many of the failure events can be explained by considering the applicability domain of the ensemble model.  In previous work we implemented transfer-learning techniques to address such limitations for gamma-ray data~\cite{moore2020transfer}.  A more comprehensive study of applicability domain and interpretable and/or explainable artificial intelligence for gamma-ray spectroscopy, including investigating self-explaining AI~\cite{Elton_2020_selfEx} is planned for future work.

\subsection{Probability and confidence}

As was noted in the sections above, simply making assignments based on which isotope class receives the highest probability value may be somewhat incomplete; during the model testing phase, many of the incorrect model assignments were made on the basis of relatively low probability values, and in many cases the correct isotope class received a non-negligible probability value.  A cursory review of the results obtained here seemed to reinforce this perspective: many of the rare isotope class assignments appear to be incorrect, but these are usually given low probability values.  

\begin{figure}
	\centering
	\hspace*{-.75in}
	\includegraphics[width=5.in]{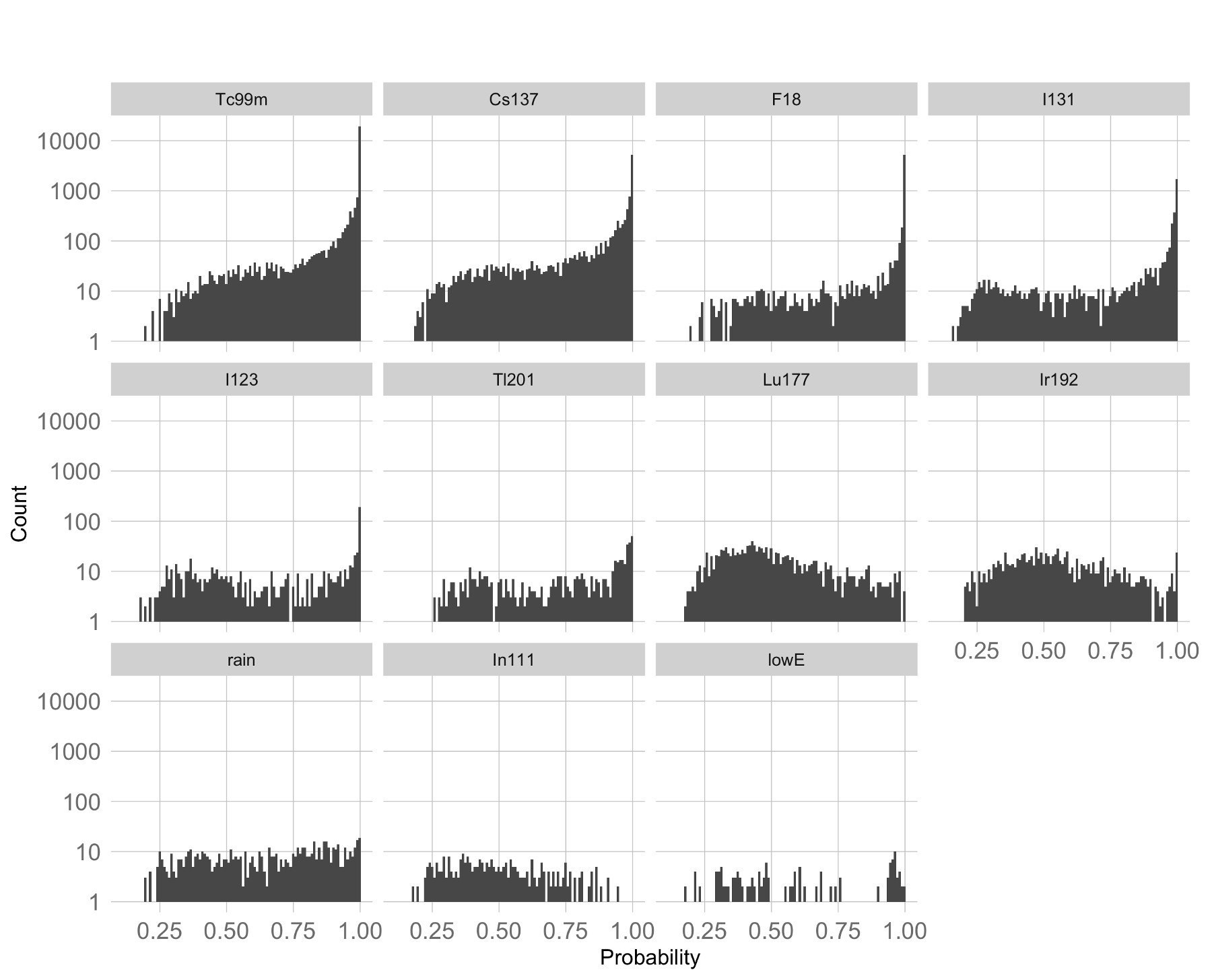}
	\caption{The distribution of probability values determined from the ensemble model for each isotope class.  The "rare isotopes" tended to yield lower probability values, whereas common isotope classes like Tc-99m yielded overwhelmingly high probability values.  These values can be used to help assess the relative confidence in the model assignments.}
	\label{fig:probability}
\end{figure}

Figure~\ref{fig:probability} shows the probability values obtained from the ensemble model for events assigned to each isotope class.  Here, the probability represents the proportion of “votes” for the given isotope class in the ensemble Random Forest model configuration~\cite{Breiman02}.  Note that there has been no effort to calibrate the probability values reported by the ensemble model or the individual model configurations that feed into it, so the term “probability” is used somewhat loosely here.  In our previous work we observed that other model configurations applied to gamma spectroscopy data suffered from poor calibration, which is not uncommon among such models~\cite{Calibration}. 

Ninety percent of the anomalies are assigned to one of the “big four” isotope classes: Tc-99m, Cs-137, F-18, and I-131, consistent with what was observed in the first 30 days.  Also, 55\% are assigned confidence values of 1.00 from the ensemble model.  From Figure~\ref{fig:probability} it is clear that the vast majority of these are associated with the “big four”.  I-123 has a significant number of “very confident” assignments, but in general the rare isotope classes exhibit a relatively flat distribution of probabilities.

\section{Discussion}

\begin{figure}
	\centering
	\hspace*{-.5in}
	\includegraphics[width=5.5in]{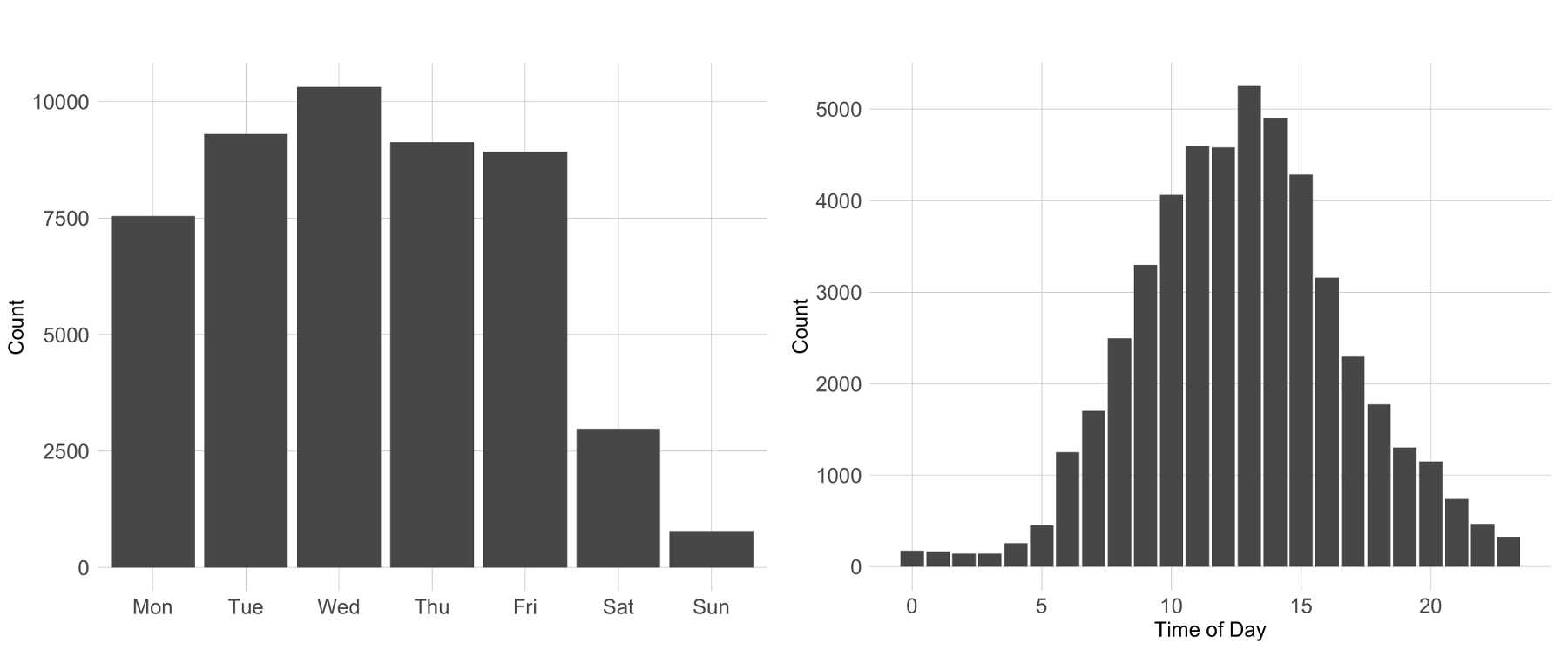}
	\caption{The total number of anomalies identified as a function of day-of-week (left) and time-of-day (right).  Each shows a distinctive pattern.}
	\label{fig:timedep}
\end{figure}

With the ensemble model results, it is possible to explore some temporal trends in the data.  For example, we can start by simply looking at the frequency of anomalies as a function of day-of-week and time-of-day, as in Figure~\ref{fig:timedep}.  Here, one can see that there are significantly fewer anomalies on the weekends as compared to weekdays.  Not only that, but a clear peak in the anomaly rate is seen just after noon, with the likelihood of anomalies between 8pm and 6am very low.  This is consistent with what one might expect since a majority of these events come from radioactive materials being transported in one way or another.

\begin{figure}
	\centering
	\hspace*{-.75in}
	\includegraphics[width=6.in]{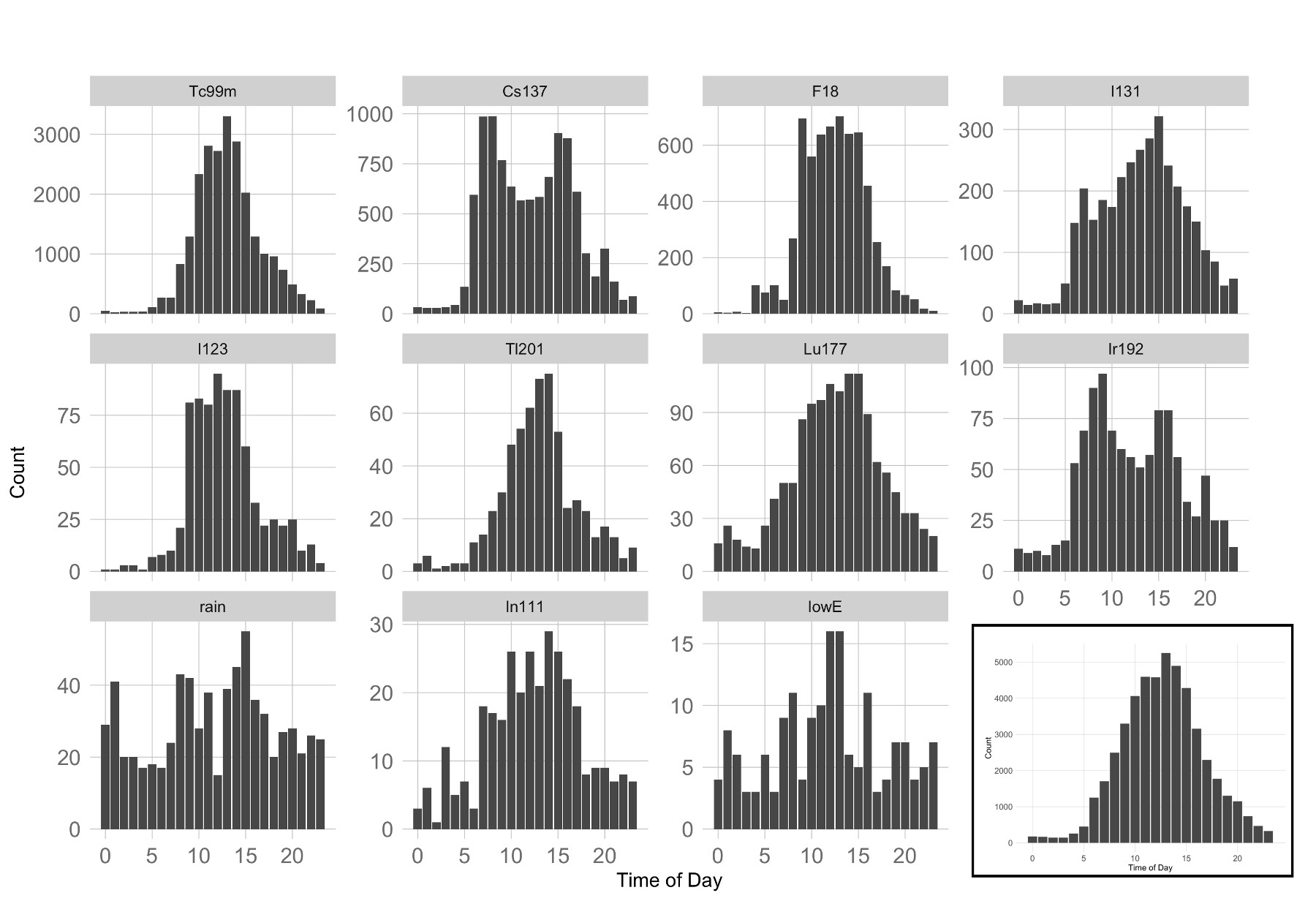}
	\caption{The distribution of each isotope class as a function of time-of-day.  For reference, the overall trend is displayed at bottom right.  Note the stark difference between profile shape exhibited by industrial radioisotopes like Cs-137 and Ir-192, and medical isotopes like Tc-99m and others.}
	\label{fig:iso_distribution}
\end{figure}

It is also possible to study these trends as a function of isotope class.  For example, in Figure~\ref{fig:iso_distribution} one can see the time-of-day results for each class.  Here, an interesting trend distinguishes common industrial isotopes like Cs-137 and Ir-192 from medical isotopes like Tc-99m, I-131, and others.  The former show a double-peaked distribution, likely the result of sources being transported to and from a work site, whereas the latter show the same distribution that could be seen in Figure~\ref{fig:timedep}, peaked just after noon.

These results are significant for a couple of reasons.  First, recall the tendency for Ir-192 anomalies to show an ambiguous shape, which we attributed to the fact that, since this radioisotope tends to be used in radiography applications, the sources are typically high in activity, but are transported in well-shielded containers.  This made the identification somewhat more difficult and, without ground truth information, it made it impossible to determine how well the models were performing.  The result here provides a strong indication that the results are reliably good for this isotope class.  It is also significant in the sense that the model is performing reasonably well in a regime where the human analyst may have difficulty.

Another isotope class worth discussing in this context is I-131.  This is a well-known medical isotope used as a radiopharmaceutical, however the profile exhibited in Figure~\ref{fig:iso_distribution} appears to be a hybrid between what we determined to be the “medical” profile and the double-peaked “industrial” profile.  At first, one might suspect that this is a sign that the model performed poorly for this isotope class, however this does not appear to be supported with a visual review of the results, and the trend is reinforced when we inspect only results reported with high confidence (model probability $>$ 0.99).  Instead, the distinct difference is likely a consequence of the fact that I-131 has a longer half-life than most of the other medical isotopes seen here, so it remains detectable by the NOVArray sensors for several days.  Thus, we see a peak in the I-131 rate in the hours after noon as with the other medical isotopes, but we also see a peak associated with the morning commute as we saw with the industrial isotopes.  By the same argument, one might expect Lu-177 to show a similar pattern, since this isotope also has a reasonably long half-life (6.7 d).  However, the histogram for this isotope class is more ambiguous and it is difficult to visually ascertain whether a similar pattern is present or not.  This may be due to poor statistics for this isotope class or, more likely, it may be due to the shorter biological half-life for this species (ie: it is flushed from the body more quickly)\cite{LU177}.

It is interesting to compare the detection frequency for each medical isotope with the frequency of use for each.  Table 2 shows the percent of radiopharmaceutical procedures that used each medical isotope~\cite{Kouzes11}, and the percent of anomalies that were identified as belonging to that isotope class (percent in the table includes only the medical isotope classes: Tc-99m, F-18, I-131, I-123, In-111, Lu-177, Tl-201).  There are some intriguing discrepancies in this table that help underscore the complexity of this comparison.  First, it was already noted above that the relatively long half-life (physical and biological) for I-131 makes it detectable for a longer period of time than the other medical isotopes.  This explains why the fraction of anomalies is so much higher than the fraction of procedures in the table.  On the other hand, F-18 has a very short half-life but makes up a notably larger percent of the anomalies than the frequency of procedures would imply.  This, and other discrepancies may actually have something to do with location-specific factors.  For example, one of the NOVArray sensors was positioned in relatively close proximity to a medical facility where radiopharmaceutical treatments are carried out, so the distributions are probably more a reflection of radioisotope use at this facility than that of the national average over all medical facilities.  

Several of the radiopharmaceuticals were not identified in the NOVArray dataset.  Some of these are fairly straightforward to explain.  For example, C-14, Sr-89, and Cr-51 do not have a significant gamma-ray signal and, therefore, wouldn’t be detected with the NOVArray sensors.  The others may not be used locally with a high frequency and therefore may not have been present (or correctly identified) in the initial 30 days dataset, or their medical use may be fundamentally different; i.e.: inpatient treatment by injection and subsequent removal of localized source pellet.  

\begin{table}\centering
	\begin{tabular}{c c c c c}
		\hline
		Isotope & Physical $t_{1/2}^{(i)}$ & Effective $t_{1/2}$ &	\% of Procedures $^{(ii)}$ & \% of Medical anomalies \\
		\hline \hline 
		Tc-99m & 6 h  & 4 h & 91.51 & 65.5 \\
		F-18 & 110 m  & 110 m & 2.02 & 17.0 \\
		C-14 & 5730 y & 40 d & 1.89 & \\
		In-111 & 2.8 h & 2.8 d & 1.40 & 0.9 \\ 
		Co-57 & 271.8 d & 9.3 d $^{(i)}$ & 1.07 & \\
		Tl-201 & 73.1 h & 72 h & 0.42 & 1.6 \\ 
		Sm-153 & 46.27 h &  & 0.39 & \\ 	
		Sr-89 & 50.5 d & 50.5 d & 0.39 & \\
		Cr-51 & 27.7 d & 22.8 d & 0.35 & \\
		I-131 &	8 d & 7.6 d & 0.34 & 9.2 \\
		Ga-67 & 3.26 d & 3.3 d & 0.19 & \\
		I-123 & 13.2 h & 13 h & 0.03 & 2.1 \\
		Lu-177 & 6.71 d & 40 h $^{(iv)}$ & N/A & 3.6 \\
		\hline
	\end{tabular}
	\caption{Relative frequency of use for medical radioisotopes and anomalies associated with medical isotopes as measured with NOVArray.
		References: $(i)$:~\cite{Grupen10}, $(ii)$:~\cite{Kouzes11}, $(iv)$:~\cite{Hosono2018}}
	\label{tab:RelFreq}
\end{table}

\section{Conclusion}

Over the course of this project, 168,336 hours of data were collected with NOVArray, and over this time 48,974 anomalies were recorded.  This gives 0.29 anomalies per detector hour, or nearly seven anomalies per day per sensor.  Using the ensemble model developed here, approximately 2/3 of those could be identified with probability greater than 0.95, and only about one anomaly per day is not one of the “big four” isotope classes (Tc99m, Cs137, F18, I131).  These values provide a fairly good idea for “what’s out there”, with the caveat that one should expect that there is a degree of geographical specificity to the results.  Nevertheless, these results provide a solid foundation for what one should expect to encounter in a radiation search.

One of the primary goals for this project was to collect a dataset for use in data analytics and machine-learning applications.  As such, the data collected over the course of this project has been made available through the Berkeley Data Service (\href{https://bdc.lbl.gov}{https://bdc.lbl.gov} ).

This work was done by Mission Support and Test Services, LLC, under Contract No. DE-NA0003624 with the U.S. Department of Energy and supported by the Site-Directed Research and Development Program. DOE/NV/03624\texttt{-{}-}1029. The United States Government retains and the publisher, by accepting the article for publication, acknowledges that the United States Government retains a non-exclusive, paid-up, irrevocable, world-wide license to publish or reproduce the published form of this manuscript, or allow others to do so, for United States Government purposes.

\bibliography{NOVArray}{}

\begin{thebibliography}{10}

\bibitem{UNEP16}
Radiation: Effects and sources, 2016.
\newblock URL:
  \url{https://www.unscear.org/unscear/en/publications/booklet.html?print}.

\bibitem{ICRU53}
Gamma spectrometry in the environment; icru report 53.
\newblock {\em ICRU}, os-27, 1994.
\newblock URL: \url{https://journals.sagepub.com/toc/crub/os-27/2}.

\bibitem{NCRP93}
Ionizing radiation exposure of the population of the united states (1987),
  1987.

\bibitem{NCRP160}
Ionizing radiation exposure of the population of the united states (2006),
  2006.

\bibitem{Jackson04}
Jackson.
\newblock Radioisotope gauges for industrial process measurements.
\newblock 2004.
\newblock \href {https://doi.org/10.1002/0470021098}
  {\path{doi:10.1002/0470021098}}.

\bibitem{NAP08}
National~Research Council.
\newblock {\em Radiation Source Use and Replacement: Abbreviated Version}.
\newblock The National Academies Press, Washington, DC, 2008.
\newblock URL:
  \url{https://www.nap.edu/catalog/11976/radiation-source-use-and-replacement-abbreviated-version},
  \href {https://doi.org/10.17226/11976} {\path{doi:10.17226/11976}}.

\bibitem{Kouzes11}
Richard {Kouzes}.
\newblock {Radiation Detection at Borders for Homeland Security}.
\newblock In {\em APS April Meeting Abstracts}, volume 2004 of {\em APS Meeting
  Abstracts}, page L2.002, May 2004.

\bibitem{OLMOS1992167}
P.~Olmos, J.C. Diaz, J.M. Perez, G.~Garcia-Belmonte, P.~Gomez, and V.~Rodellar.
\newblock Application of neural network techniques in gamma spectroscopy.
\newblock {\em Nuclear Instruments and Methods in Physics Research Section A:
  Accelerators, Spectrometers, Detectors and Associated Equipment}, 312(1):167
  -- 173, 1992.
\newblock URL:
  \url{http://www.sciencedirect.com/science/article/pii/016890029290148W},
  \href {https://doi.org/https://doi.org/10.1016/0168-9002(92)90148-W}
  {\path{doi:https://doi.org/10.1016/0168-9002(92)90148-W}}.

\bibitem{Portnoy04}
David Portnoy, Peter Bock, Peter~C. Heimberg, and Eric~T. Moore.
\newblock {Using ALISA for high-speed classification of the components and
  their concentrations in mixtures of radioisotopes}.
\newblock In F.~Patrick Doty, Richard~C. Schirato, H.~Bradford Barber, and Hans
  Roehrig, editors, {\em Penetrating Radiation Systems and Applications VI},
  volume 5541, pages 1 -- 10. International Society for Optics and Photonics,
  SPIE, 2004.
\newblock \href {https://doi.org/10.1117/12.555833}
  {\path{doi:10.1117/12.555833}}.

\bibitem{Sharma12}
S.~{Sharma}, C.~{Bellinger}, N.~{Japkowicz}, R.~{Berg}, and K.~{Ungar}.
\newblock Anomaly detection in gamma ray spectra: A machine learning
  perspective.
\newblock In {\em 2012 IEEE Symposium on Computational Intelligence for
  Security and Defence Applications}, pages 1--8, July 2012.
\newblock \href {https://doi.org/10.1109/CISDA.2012.6291535}
  {\path{doi:10.1109/CISDA.2012.6291535}}.

\bibitem{Gerrit17}
Marc~Gerrit Paff, Angela {Di Fulvio}, Shaun~D. Clarke, and Sara~A. Pozzi.
\newblock Radionuclide identification algorithm for organic scintillator-based
  radiation portal monitor.
\newblock {\em Nuclear Instruments and Methods in Physics Research Section A:
  Accelerators, Spectrometers, Detectors and Associated Equipment}, 849:41 --
  48, 2017.
\newblock URL:
  \url{http://www.sciencedirect.com/science/article/pii/S0168900217300074},
  \href {https://doi.org/https://doi.org/10.1016/j.nima.2017.01.009}
  {\path{doi:https://doi.org/10.1016/j.nima.2017.01.009}}.

\bibitem{Kamuda2017}
Mark Kamuda, Jacob Stinnett, and Clair Sullivan.
\newblock Automated isotope identification algorithm using artificial neural
  networks.
\newblock {\em IEEE Transactions on Nuclear Science}, 64(7), 4 2017.
\newblock \href {https://doi.org/10.1109/TNS.2017.2693152}
  {\path{doi:10.1109/TNS.2017.2693152}}.

\bibitem{Masala19}
Eugene Masala and Laura Blomeley.
\newblock Machine-learning algorithm for shielded special nuclear materials
  detection.
\newblock {\em CNL Nuclear Review}, 8(2):145--157, 2019.
\newblock \href {http://arxiv.org/abs/https://doi.org/10.12943/CNR.2018.00004}
  {\path{arXiv:https://doi.org/10.12943/CNR.2018.00004}}, \href
  {https://doi.org/10.12943/CNR.2018.00004}
  {\path{doi:10.12943/CNR.2018.00004}}.

\bibitem{Durbins2020}
Matthews Durbin and Azaree Lintereur.
\newblock Implementation of machine learning algorithms for detecting missing
  radioactive material.
\newblock {\em Journal of Radioanalytical and Nuclear Chemistry},
  324(3):1455--1461, 2020.
\newblock \href {https://doi.org/10.1007/s10967-020-07188-4}
  {\path{doi:10.1007/s10967-020-07188-4}}.

\bibitem{ford2019threat}
William~P. Ford, Emma Hague, Tom McCullough, Eric Moore, and Johanna Turk.
\newblock Threat determination for radiation detection from the remote sensing
  laboratory, 2019.
\newblock \href {http://arxiv.org/abs/1908.11207} {\path{arXiv:1908.11207}}.

\bibitem{moore2019application}
Eric~T. Moore, William~P. Ford, Emma~J. Hague, and Johanna Turk.
\newblock An application of cnns to time sequenced one dimensional data in
  radiation detection, 2019.
\newblock \href {http://arxiv.org/abs/1908.10887} {\path{arXiv:1908.10887}}.

\bibitem{Hague_2019}
Emma~J. Hague, Mark Kamuda, William~P. Ford, Eric~T. Moore, and Johanna Turk.
\newblock A comparison of adaptive and template matching techniques for
  radio-isotope identification.
\newblock {\em Algorithms, Technologies, and Applications for Multispectral and
  Hyperspectral Imagery XXV}, May 2019.
\newblock URL: \url{http://dx.doi.org/10.1117/12.2519062}, \href
  {https://doi.org/10.1117/12.2519062} {\path{doi:10.1117/12.2519062}}.

\bibitem{moore2020transfer}
Eric~T. Moore, Johanna~L. Turk, William~P. Ford, Nathan~J. Hoteling, and
  Lance~S. McLean.
\newblock Transfer learning in automated gamma spectral identification, 2020.
\newblock \href {http://arxiv.org/abs/2003.10524} {\path{arXiv:2003.10524}}.

\bibitem{VA2020}
Virginia department of transportation.
\newblock \url{https://www.virginiaroads.org}.
\newblock Accessed: 2020-11-14.

\bibitem{Anom2020}
Private communication.

\bibitem{Duncan98}
Kara Duncan.
\newblock Radiopharmaceuticals in pet imaging.
\newblock {\em J. Nucl. Med. Technol.}, 26(4):228--234, 1998.

\bibitem{unbalanced}
Bartosz Krawczyk.
\newblock Learning from imbalanced data: open challenges and future directions.
\newblock {\em Progress in Artificial Intelligence}, 5, 2016.
\newblock \href {https://doi.org/10.1007/s13748-016-0094-0}
  {\path{doi:10.1007/s13748-016-0094-0}}.

\bibitem{Johnson19}
Justin~M. Johnson and Taghi~M. Khoshgoftaar.
\newblock Survey on deep learning with class imbalance.
\newblock {\em Journal of Big Data}, 2019.
\newblock \href {https://doi.org/10.1186/s40537-019-0192-5}
  {\path{doi:10.1186/s40537-019-0192-5}}.

\bibitem{Rproject}
The r project for statistical computing.
\newblock \url{https://www.r-project.org/}.
\newblock Accessed: 2020-11-23.

\bibitem{appDomainQSAR}
F.~Sahigara, K.~Mansouri, D.~Ballabio, A.~Mauri, V.~Consonni, and
  R.~Todeschini.
\newblock Comparison of different approaches to define the applicability domain
  of qsar models.
\newblock {\em Molecules}, 17:4791–4810, 2012.
\newblock \href {https://doi.org/10.3390/molecules17054791}
  {\path{doi:10.3390/molecules17054791}}.

\bibitem{Elton_2020_selfEx}
Daniel~C. Elton.
\newblock Self-explaining ai as an alternative to interpretable ai.
\newblock {\em Lecture Notes in Computer Science}, page 95–106, 2020.
\newblock URL: \url{http://dx.doi.org/10.1007/978-3-030-52152-3_10}, \href
  {https://doi.org/10.1007/978-3-030-52152-3_10}
  {\path{doi:10.1007/978-3-030-52152-3_10}}.

\bibitem{Breiman02}
LEO BREIMAN.
\newblock Random forests.
\newblock {\em Machine Learning}, 2001.

\bibitem{Calibration}
Chuan Guo, Geoff Pleiss, Yu~Sun, and Kilian~Q. Weinberger.
\newblock On calibration of modern neural networks.
\newblock {\em CoRR}, abs/1706.04599, 2017.
\newblock URL: \url{http://arxiv.org/abs/1706.04599}, \href
  {http://arxiv.org/abs/1706.04599} {\path{arXiv:1706.04599}}.

\bibitem{LU177}
Makoto Hosono, Hideharu Ikebuchi, Yoshihide Nakamura, Nobutaka Nakamura,
  Takahiro Yamada, Sachiko Yanagida, Asami Kitaoka, Kiyotaka Kojima, Hiroyasu
  Sugano, Seigo Kinuya, Tomio Inoue, and Jun Hatazawa.
\newblock Manual on the proper use of lutetium-177-labeled somatostatin
  analogue (lu-177-dota-tate) injectable in radionuclide therapy (2nd ed.).
\newblock {\em Annals of nuclear medicine}, 32:217--235, 2018.
\newblock URL: \url{https://pubmed.ncbi.nlm.nih.gov/29333565}, \href
  {https://doi.org/10.1007/s12149-018-1230-7}
  {\path{doi:10.1007/s12149-018-1230-7}}.

\bibitem{Grupen10}
Claus Grupen.
\newblock {\em Introduction to Radiation Protection}.
\newblock 2010.

\bibitem{Hosono2018}
Makoto Hosono, Hideharu Ikebuchi, Yoshihide Nakamura, Nobutaka Nakamura,
  Takahiro Yamada, Sachiko Yanagida, Asami Kitaoka, Kiyotaka Kojima, Hiroyasu
  Sugano, Seigo Kinuya, Tomio Inoue, and Jun Hatazawa.
\newblock Manual on the proper use of lutetium-177-labeled somatostatin
  analogue (lu-177-dota-tate) injectable in radionuclide therapy (2nd ed.).
\newblock {\em Annals of Nuclear Medicine}, 32:217 -- 235, 2018.
\newblock \href {https://doi.org/10.1007/s12149-018-1230-7}
  {\path{doi:10.1007/s12149-018-1230-7}}.

\end{thebibliography}
\bibliographystyle{unsrturl} 

\end{document}